\documentclass[%
 reprint,
 amsmath,amssymb,
 aps,
pre,
longbibliography]{revtex4-1}

\usepackage{graphicx}
\usepackage{dcolumn}
\usepackage{bm}
\usepackage{natbib}
\usepackage{siunitx}
\usepackage{amsmath}
\usepackage[utf8]{inputenc}
\usepackage{mathtools}
\usepackage{wasysym}
\usepackage{xcolor}
\usepackage{hyperref}
\hypersetup{
    colorlinks=true,
    linkcolor=blue,
    filecolor=magenta,      
    urlcolor=magenta,
}
\usepackage{marvosym}

\makeatletter
\def\@bibdataout@aps{%
\immediate\write\@bibdataout{%
@CONTROL{%
apsrev41Control%
\longbibliography@sw{%
    ,author="08",editor="1",pages="1",title="0",year="1"%
    }{%
    ,author="08",editor="1",pages="1",title="",year="1"%
    }%
  }%
}%
\if@filesw \immediate \write \@auxout {\string \citation {apsrev41Control}}\fi 
}
\makeatother

\begin{document}

\preprint{AIP/ApplRevLett}

\title{Conformally mapped black hole effect in elastic curved continuum}

\author{Dongwoo Lee$^{1}$}%
 \email{These authors contributed equally to this work.}
\author{Yiran Hao$^{2}$}%
 \email{These authors contributed equally to this work.}
\author{Jeonghoon Park$^{1}$}%
\author{Yaxi Shen$^{2}$}%
\author{Jensen Li$^{2}$}%
 \email{jensenli@ust.hk}
\author{Junsuk Rho$^{1,3,4}$}%
 \email{jsrho@postech.ac.kr}
 
\affiliation{%
$^{1}$Department of Mechanical Engineering, Pohang University of Science and Technology (POSTECH), Pohang 37673, Republic of Korea
\\$^{2}$Department of Physics, The Hong Kong University of Science and Technology, Clear Water Bay, Kowloon, Hong Kong, China
\\$^{3}$Department of Chemical Engineering, Pohang University of Science and Technology (POSTECH), Pohang 37673, Republic of Korea
\\$^{4}$POSCO-POSTECH-RIST Convergence Research Center for Flat Optics and Metaphotonics, Pohang 37673, Republic of Korea
}%

\date{\today}

\begin{abstract}
We present a black hole effect by strategically leveraging a conformal mapping in elastic continuum with curved-space framework, which is less stringent compared to a Schwarzschild model transformed to isotropic refractive index profiles. In the conformal map approach, the 2D point singularity associated to the black hole effect is accomplished by physical plates with near-to-zero thickness. The analog gravity around the singularity results in highly confined energy and lagged timings within a branch cut of the conformal map. These effects are quantified both numerically and experimentally in reference to control trials in which the thickness is not modulated. The findings would deepen our understanding of the elastic analog in mimicking gravitational phenomena, as well as establish the elastic continuum framework for developing a generic design recipe in the presence of the index singularity. Geometric landscapes with elastically curved surfaces would be applicable in a variety of applications such as sensing, imaging, vibration isolation, and energy harvesting.   

\begin{description}
\item[Usage]
Accepted.
\end{description}
\end{abstract}

\pacs{Valid PACS appear here}
\maketitle

\renewcommand{\vec}[1]{\mathbf{#1}}
\section{\label{sec:level1}Introduction}
A black hole is an enigmatic cosmic phenomenon with an endless curvature and a singularity \cite{penrose1,penrose2}. Since the discovery of gravitational waves from binary black holes \cite{binaryBH}, there has been a rising interest in questioning the feasibility of equivalent phenomena in classical waves. It is very difficult to envisage or construct black holes in classical laboratory frame models due to the stringent constraint on massive mass (i.e., space-time distortion) with the interplay between space-time and energy-momentum (matter) information in general relativity \cite{relativity}. Alternatively, we pursue curved elastic continuum to reveal the distortion equivalent to refractive index with spatial curvature for the emergence of such a black hole.

Great efforts have been made in the metamaterial arena to supply effective parameters on each of the macroscopic unit cells relating to unprecedented wave-bending capabilities. Metamaterial building blocks, for example, have been used to illustrate optical black holes \cite{opticalBH4,opticalBH1,opticalBH3,opticalBH2}, wormholes \cite{opticalWH}, and metric signature transitions \cite{metricsignature}. More recently, geometric modulated landscapes have shown their versatility when employed in elastic versions of wormholes \cite{elasticWH}, invisibility cloaks \cite{elasticcloak}, and gravitational lensing effects \cite{elasticlensing} without the usage of resonating components that could be a nuisance due to their lossy nature and narrow working frequency range. Given the presence of richer polarization states in underlying elasticity, as is the case in other metamaterial mainstream areas, one may take use of elastic continuum models by controlling the bending stiffness or mass density of the desired beams and rods to obtain extremely large indices \cite{continuum1,continuum2,continuum3}.

Interestingly, the notion of an acoustic black hole (ABH) \cite{mironov} has been adopted and recently investigated using a wedge plate for both straight \cite{flatBH1,flatBH2,flatBH3} and spiral \cite{spiralBH1,spiralBH2} designs for elastic wave dampening. However, the current strategy is heavily reliant on a power-law profile of plate thickness ($h(x)=\varepsilon x^m$ for $m \geq 2$). This is unrelated to transformation optics \cite{TOoptics1,TOoptics2} or acoustics \cite{TOacoustics}, space-time formulae \cite{cosmologyTransform}, or conformal mapping \cite{conformal1,conformal2}. In this respect, when extended to two-dimensional (2D) and three-dimensional (3D) space, the profile and ray dynamics are still constrained to the one-dimensional (1D) scenario. Moreover, in terms of preassigning $m$ to seek the best optimized $\varepsilon$ and $m$ with growing spatial length, a trial-and-error method is unavoidable.

In this work, we describe an innovative stance to implementing curved space in the similar way as lensing effects work in determining the refractive index. By comparing a previous Minkowski metric black hole formulation \cite{opticalBH4,cosmologyTransform} and a conformal mapping approach, we explore two index singularity models. It turns out that the former has spinning rays due to photon sphere effect and background index mismatching, while the latter has nonspinning rays that head toward the singularity. Therefore, we leverage conformal mapping to engage the warped coordinate information to the elastic curved continuum. Renowned ramifications for black holes, such as wave condensation and prolonged durations within the branch cut which plays an analog role in the event horizon of a black hole, are observed numerically and experimentally with good agreement.


\section{\label{sec:level1}Ray trajectories and models}
We compare two black hole effects by using a 3+1 pseudo-Euclidean space (Minkowski space-time) and a conformal mapping, respectively. The former is typically referred to the model by Schwarzschild \cite{schwarzschild}. The generic four-metric form including time can be expressed as \cite{opticalBH4,cosmologyTransform}
\begin{equation} \label{eq_1}
ds^2=-g_{00}(r) c^2 dt^2+g_r(r)dr^2+r^2 d\Omega^2,
\end{equation}
where $g_r(r)=g^{-1}_{00}(r)$ and $\Omega^2=d\theta^2+\sin^2{\theta}d\phi^2$ with the line element $s$. $c$, the speed of rays, is set to a constant value of 1 for convenience. By mapping into the radial direction with a variable $R$ with respect to a radius in spherical coordinate to be transformed, we obtain
\begin{equation} \label{eq_2}
ds^2=-g_{00}(r(R))dt^2+g(r(R))(dR^2+R^2 d\Omega^2),
\end{equation}
where $g(r(R))=r^2/R^2=g_r(r(R))dr^2/dR^2$. Then we have an isotropic refractive-index formula as $n(R)=(g/g_{00})^{1/2}=|dr/dR|/g_{00}(r(R))=r/(R\sqrt{g_{00}(r(R))})$. Considering for the Schwarzschild black hole, $g_{00}(r)=1-r_{eh}/r$ holding $r\gg r_{EH}$ where $r_{EH}$ is the radius from the singularity to the event horizon. This scenario supports that waves or rays cannot escape once they come inside the event horizon. To obtain the analytic expression of refractive index, $R$ and $r$ are solved as follows:
\begin{equation} \label{eq_3}
\begin{split}
R(r) & =\exp{[2\tanh^{-1}{\Big[\sqrt{1-\frac{r_{EH}}{r}}\Big]}]}, \\
  & = \frac{2r-r_{EH}+2\sqrt{r^2-r_{EH}r}}{r_{EH}},
\end{split}
\end{equation}
where $r\in[r_{EH},\infty]$ and $R\in[r_{EH}/4,\infty]$. Then, $r$ is obtained solving such a condition:
\begin{equation} \label{eq_4}
r(R)=\frac{r_{EH}(1+R)^2}{4R},
\end{equation}
where it gives rise to the isotropic refractive index for realization \cite{cosmologyTransform} assuming the parameters $r_{EH}$ and $R$ belong to real numbers to avoid complex values for convenience:
\begin{equation} \label{eq_5}
n(R)=\frac{r_{EH}|R+1|^3}{4 R^2|R-1|}.
\end{equation}

\begin{figure}[htp]
\includegraphics [width=0.47\textwidth] {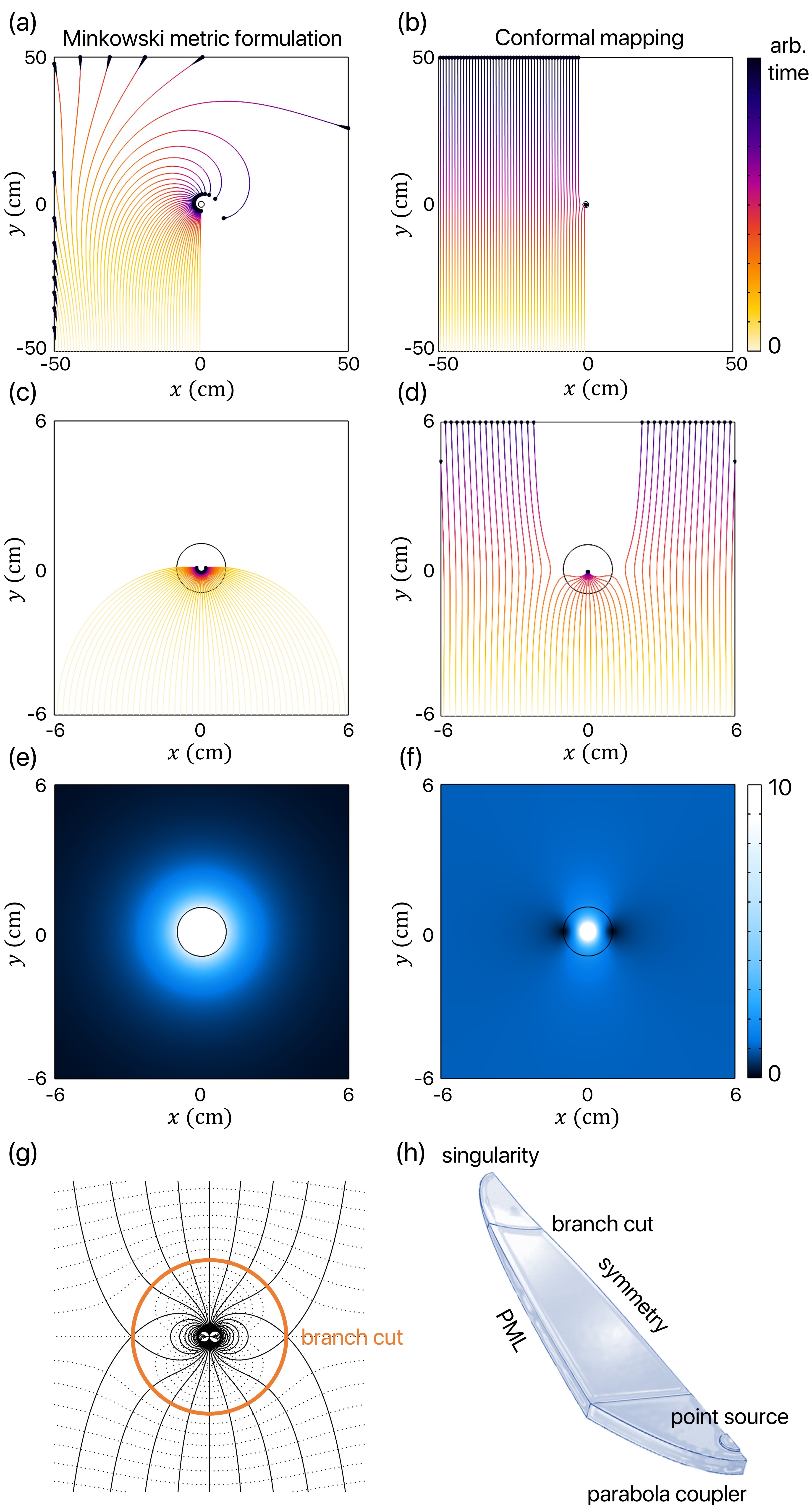}
\caption{Ray trajectories by (a) and (c) Minkowski metric formulation, and (b) and (d) conformal mapping. When rays are stimulated far out from the event horizon, spinning rays arise in the Schwarzschild model owing to the photon sphere effect in the upper panel. In the conformal strategy, no spinning rays are observed. The lower panel illustrates that when rays are excited near the event horizon or branch cut, they all end up at the singularity. It is worth noting that we choose timeframes at random with the arbitrary spatial dimensions and the fixed radius $R=1$ cm (black circle). Index distributions are also calculated for (e) the Schwarzschild model and (f) the confomal mapping. (g) Conformal map. Black solid and dotted lines denote real-valued and imaginary-valued $\zeta$, respectively. Orange line ($a=1$ cm) indicates the branch cut in $z$ space. (h) Final design of an elastic analog continuum in which the deformed coordinate is replaced by a curved tangible structure.}
\label{fig1}
\end{figure}

In the conformal mapping strategy, on the other hand, two complex planes can be routed to define the refractive index in physical space as
\begin{equation} \label{eq_6}
n = n'\Big|\frac{d\zeta}{dz}\Big|,
\end{equation}
where $n'$ is the refractive index in virtual space where the original coordinate that corresponds to $z=x+iy$ is considered. By employing the Joukowsky mapping
\begin{equation} \label{eq_7}
\zeta(z)=z+a^2/z,
\end{equation}
or equivalently $z(\zeta)=\frac{\zeta \pm \sqrt{\zeta^2-4 a^2}}{2}$ with branch points $\pm 2a$. Eq. \ref{eq_7} results in the transformed coordinate, and we can obtain a black hole absorption effect by deliberately picking the region with real-valued $\zeta$ throughout the entire map while the branch cut $a$ and the singularity at the origin $(0,0)$ are present. Following that, the index profile is formulated as $n=n'|1- a^2/z^2|$ where $n'=1$. We quantify ray trajectories for both approaches in Figs.~\ref{fig1}(a-d) 
and reveal that spinning rays excited far from the event horizon emerge in the Minkowski metric formulation due to the photon sphere effect, particularly when an isotropic refractive index profile is made; however, well-localized rays towards the singularity can be obtained using conformal transformation. Moreover, for the Minkowski model near the event horizon, the rays do not spin but instead exhibit background index mismatching ($n<1$) [Fig.~\ref{fig1}(e)] that relaxes the index range in the conformal strategy [Fig.~\ref{fig1}(f)], resulting in a complicated structure. These findings inspire us to use the conformal mapping method for realization and to focus on the enlarged area surrounding the branch cut. Note that the term \textit{spinning} does not refer to the Kerr black hole's rotating feature along its axis symmetry. In Fig.~\ref{fig1}(g), the conformal map corresponding to the upper sheet of the Riemann surface that we can perceive only in physical space is calculated, and the orange circle symbolizes a branch cut in $z$ space, which is a complex plane curve that crosses the discontinuous boundary of an analytic multivalued function between branch points $\pm 2a$. The black solid and dotted lines indicate Re$(\zeta)$ and Im$(\zeta)$, respectively. We can observe in this model that rays radiating from either the top or bottom sides uniformly reach the central singularity. Hereinafter, we simply call the absorption effect based on singularity as the black hole absorption effect. We note that the conformal map \cite{conformal1}, in fact, provides the index range $n\in[1,\infty]$, which explains why practical implementation did not happen 20 years ago. We may still explore the effects using our current method if we continue to use a nonsingular index value at the center. Therefore, we make use of the conformal map for Re$(\zeta)$ region in order to reflect it into the corresponding physical plate model [Fig.~\ref{fig1}(h)]. The refractive index in deformed space is in analogy to the thickness of plates in elastic continuum. To characterize refractive index, we use the Kirchhoff-Love plate theory $(D \nabla^4 u+2\rho h (\partial^2 u/\partial t^2) = 0)$ where $D$ is the bending stiffness, $\rho$ is the mass density and $h$ is the half thickness, which describes the flexural motion $u$ (i.e., antisymmetric Lamb wave ($A_0$ mode) of thin plates in the elastic continuum side. Given monochromatic plane waves with waven umber $k$ and angular frequency $\omega$, a quadratic dispersion relation is derived as $k^4 = 2\rho h\omega^2/D$ and is satisfied in relation to the refractive-index and the phase velocity under the linear regime. It turns out that when the values of the other elastic variables are maintained constant, the refractive-index is approximately equivalent to $h^{-1/2}$. Thus, once the warped coordinate information is established, we can immediately extract the refractive index data, which maps into plate thickness as $h(x,y)=\sqrt{|1-a^2/z^2|}$. By relying only on the geometry effect, this is hugely beneficial throughout a wide range of refractive index values with broadband performance. Following the relationship, we design the elastic black hole and cut along the intersection of the whole map and the boundary specified by a quadratic function $y=-(L/2 l)^2 x^2$ where $(x,y):[-L,L]$, $L=18$ cm, and $l=L/5$. The schematic of the final design is illustrated in Fig.~\ref{fig1}(h), with an extra parabola coupler to excite a plane-wave-like source from a point source through parabolic geometry, and perfectly matched layer (PML) and symmetry conditions implemented on both sides for efficient numerical computation. Take note that the PML condition is applied outside the branch cut, while the remaining condition is the free boundary inside the branch cut. To confirm undiluted singularity absorption, we employ no additional damping materials in the vicinity of the singularity before comparing the damping material ingredient. The adoption of the viscoelastic damping layer clearly illustrates the lengthened duration with reduced reflection inside the branch cut, mimicking the effect of a black hole.

\section{\label{sec:level1}Frequency response}
\begin{figure}[htp]
\includegraphics [width=0.47\textwidth] {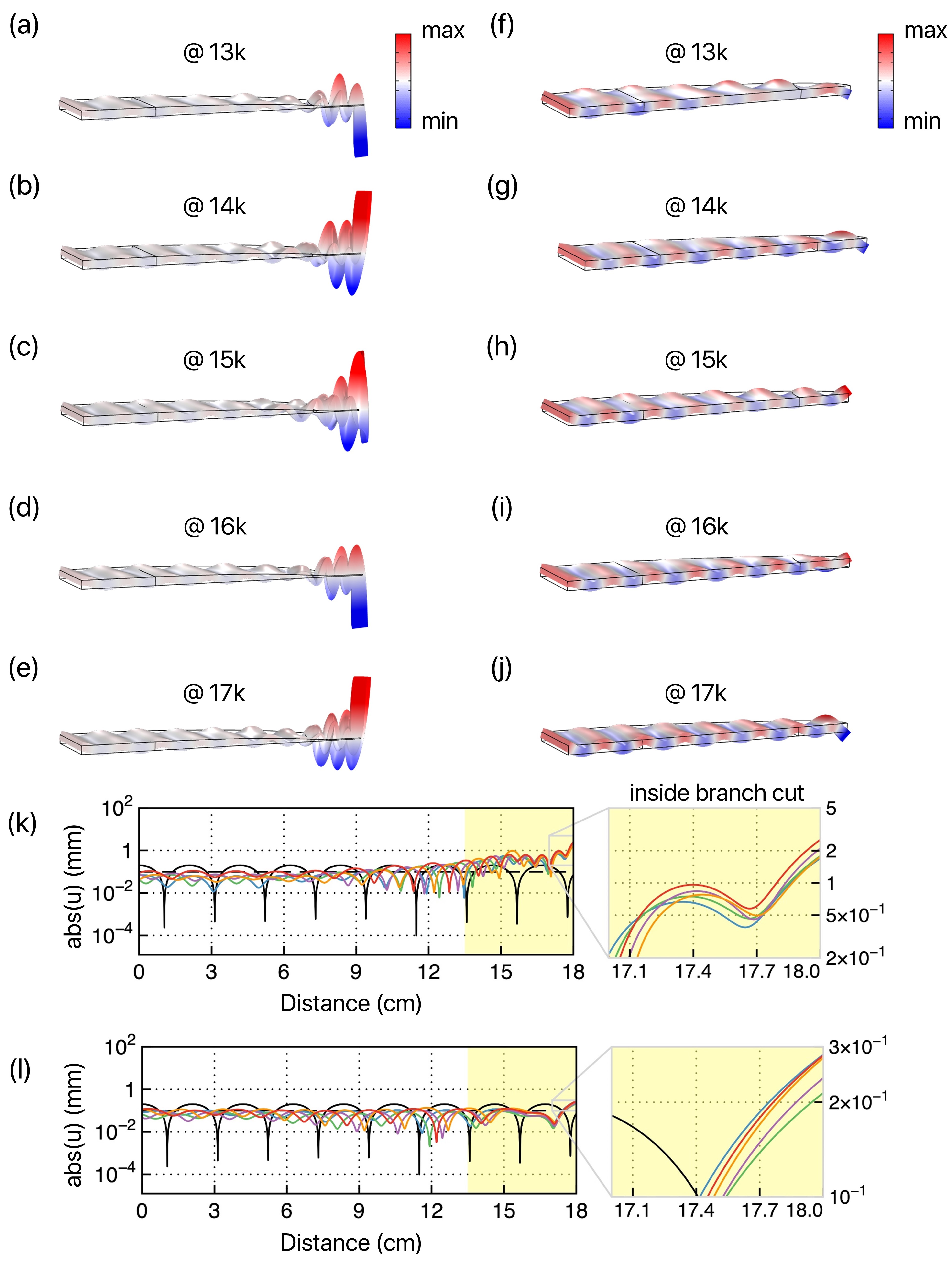}
\caption{Out-of-plane displacement fields with harmonic excitation for (a)-(e) black hole and (f)-(j) control trial models at several frequencies. (k) and (l) are the absolute values of the out-of-plane displacement fields in both examples, which are compared analytically to total reflection (black solid line) and no reflection (black dashed line), respectively. Several frequencies are as follows: 13 kHz (blue), 14 kHz (green), 15 kHz (red), 16 kHz (purple), and 17 kHz (orange). Only the black hole model is capable of producing an energy accumulation transition that exceeds the cutoff $|u|<0.2$ mm.}
\label{fig2}
\end{figure}

We investigate plane wave incidence directly by connecting a flat plate with a prescribed load $u=0.1$ mm pulsating uniformly along the $z$ axis. The maximum and minimum thicknesses are $7.26$ and $0.8$ mm that correspond to the effective index range $[1,3.0125]$, with a target frequency of $15$ kHz. The material properties are determined by $\rho=1190$ $\mathrm{kg/m^3}$, $E=3.4$ $\mathrm{GPa}$, and $\nu=0.35$. Figs.~\ref{fig2}(a)-(e) for the black hole sample highlight that highly confined out-of-plane displacement fields are observed around singularity inside the branch cut over a broad frequency range. Under the condition $\lambda_\textrm{cut on} < 2 a$ corresponding to the cut-on frequency that is a threshold \cite{gm1,gm2}, where $\lambda_\textrm{cut on}$ is the cut-on wavelength and $a=4.5$ cm, the geometry effect-dependent broadband performance becomes functional. Such a condition implies that the confined and delayed mechanism works locally by preventing global modes of the whole plate. $\lambda_\textrm{cut on}$ can be estimated by $c=(\omega^2 E/(3(1-\nu)^2\rho))^{1/4} \sqrt{h}$ for the flexural wave speed of the reference plate by equating $c=\lambda_\textrm{cut on} f$, resulting in $\lambda_\textrm{cut on} \approx 3.38$ cm at 15 kHz, which is much less than $2a$. 
The weak fields outside the branch cut reveal the tiny reflected waves caused by the minimal but still finite thickness at the singularity. A control sample with no thickness modulation, on the other hand, produces reflecting field patterns with no confined out-of-plane displacement fields [Fig.~\ref{fig2}(f)-(j)]. The black hole sample, in particular, exhibits energy condensation within the branch cut. At the singularity, the maximum absolute value of out-of-plane displacement field $|u|_{\text{max}}$ is approximately 25.3 times greater than that of the incident $|u|_{\text{in}}$ [Fig.~\ref{fig2}(k)]. This is robust by employing a 2D $n$ profile rather than a 1D profile of $n$ that was used in earlier ABHs \cite{mironov,flatBH1,flatBH2,flatBH3} within the same degree of thickness modulation. However, the control sample yields only $|u|_{\text{max}}/|u|_{\text{in}}\approx 2.8$ mainly owing to reflection by the round shape at the tip [Fig.~\ref{fig2}(l)]. It is worth mentioning that the confinement mechanism provided by the black hole effect causes a very significant deflection oscillating in the $z$ axis within the branch cut and singularity. This affects the geometric nonlinearity despite the fact that the magnitude of the described load on the source side is small relative to the plate's spatial dimensions. To this end, we perform full-wave simulations with the geometric nonlinearity included, and we find that the resulting decrease in localized efficiency for the curved plate design is $|u|_{\text{max}}/|u|_{\text{in}}\approx 13$. However, as there is no significant confinement present at the tip for the control sample, the geometric nonlinearity is indeed missing, where $|u|_{\text{max}}/|u|_{\text{in}}\approx 2.8$ is obtained in a constant way. We also compute the extent of reflection by comparing it to the analytic formula, $|u|=|u^{+}(e^{j k d}+\mathcal{R} e^{-j k d})|$, where $d=L-y$, $u^{+}=0.1$ mm, and $\mathcal{R}$ is the reflection coefficient. We find a constant $|u|$ (black dashed line) as 0.1 for $\mathcal{R}=0$ and a substantial fluctuation on $|u|$ (black solid line) between 0 and 0.2 for $\mathcal{R}=1$. The partial reflection with the fluctuation on $|u|$ in the range of $0<|u|<0.2$ manifests in both scenarios, but the energy accumulation process operates within the branch cut only in the black hole sample, indicating that the most values have $|u|\gg 0.2$. Thus, employing curved continuum, extremely confined elastic waves can be achieved at the small segment.

\section{\label{sec:level1}Temporal response}
\begin{figure}[htp]
\includegraphics [width=0.47\textwidth] {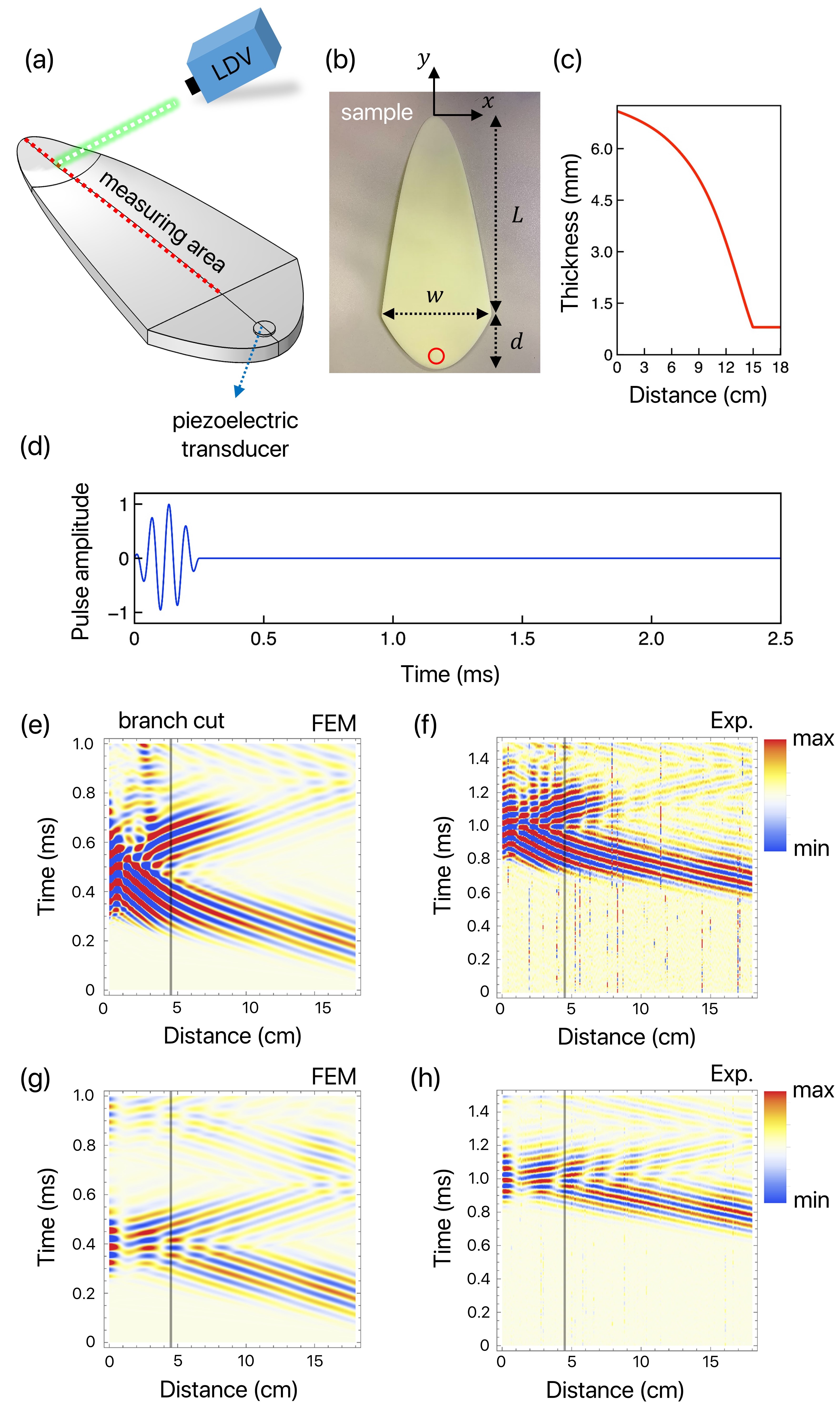}
\caption{(a) Schematic of an experimental setup that includes a piezoelectric transducer and a LDV that scans the measuring area (red dashed line) to obtain the velocity field map. (b) Fabricated sample image with spatial dimensions ($L, w, d$) that define the parabola coupler. (c) Corresponding thickness distribution along the measuring area in (a). (d) Pulse shape defined by Eq. (\ref{eq_8}) that is modulated by sine function. Spatiotemporal velocity field maps obtained by (e) FEM and (f) experiments for the black hole model, and (g) and (h) for the control trial model without thickness modulation.}
\label{fig3}
\end{figure}

We conduct time-domain experiments in obtaining field profiles in both spatial and temporal resolution. The measuring process is illustrated schematically in Fig.~\ref{fig3}(a),
which includes a piezoelectric transducer and a laser Doppler vibrometer (LDV) as a source and a sensor, respectively. The measurement area is defined by a red dotted line in the center along the $y$ axis, and the transducer is affixed to the focal point of the parabola coupler. The coupler is designed by $4P(y+L+d)=x^2$ where $d=5.0912$ cm and $w=2d$, and the focal point (red circle) is situated at $(0,-L-d+P)$ with $P=1.2728$ cm. The whole sample manufactured by stereolithography 3D printer and the thickness profile are seen in Fig.~\ref{fig3}(b) and Fig.~\ref{fig3}(c). We send a pulse and take continuous snapshots of velocity fields to evaluate the temporal response. The pulse is defined by
\begin{equation} \label{eq_8}
u_{\text{pulse}} = \sin{(2 \pi f_m t)}\times \cos{(2 \pi f_c t)} \hspace{0.2cm} \text{for}\hspace{0.2cm} t \leq 1/(2 f_m),
\end{equation}
where $u_{\text{pulse}}=0$ for $t>1/(2 f_m)$. $f_m (= 2$ kHz) and $f_c (= 15$ kHz) are the modulated and carrier frequencies, and the pulse width in time is equal to $0.25$ ms [Fig.~\ref{fig3}(d)]. Without loss of generality, we compare the black hole sample and the control sample analyzed by both finite element method (FEM) and experiment under the same conditions. The spatiotemporal velocity field map is measured, as shown in Figs.~\ref{fig3}(e)-\ref{fig3}(h), and the black vertical line denotes the branch cut. Once the elastic waves cross the branch cut, the black hole sample has a prolonged stay duration with confinement cling to the singularity [Fig.~\ref{fig3}(e)], while the control trial sample yields direct reflection at the tip [Fig.~\ref{fig3}(g)]. The FEM-calculated spatiotemporal behavior is compatible with the experimental results [Figs.~\ref{fig3}(f) and \ref{fig3}(h)]. However, in experiments, the starting time has a constant delay of approximately $0.5$ ms in practical excitation, comparing to simulation. It indicates that waves instantaneously penetrate through the branch cut when the energy accumulation is not responsible for both samples at initial times. Notably, the black hole sample clearly offers a longer stay duration even at the timings of two consecutive pulse widths in time, and the waveform indeed exemplifies that most waves dwell within the branch cut with low reflection in the meantime, mainly owing to the small residual thickness surrounding the singularity. More accumulated elastic energy can be captured around the corner when waves stay longer within the branch cut under the harmonic wave rather than a short pulse. In the flat sample, however, arriving waves immediately reflect off the tip and have already traveled back to the origin, passing through the branch cut. Figs.~\ref{fig3}(g) and \ref{fig3}(h) represent that we can no longer distinguish the confinement, and observes recurring reflection at both ends with a fluctuation in velocity fields.
\begin{figure}[htp]
\includegraphics [width=0.47\textwidth] {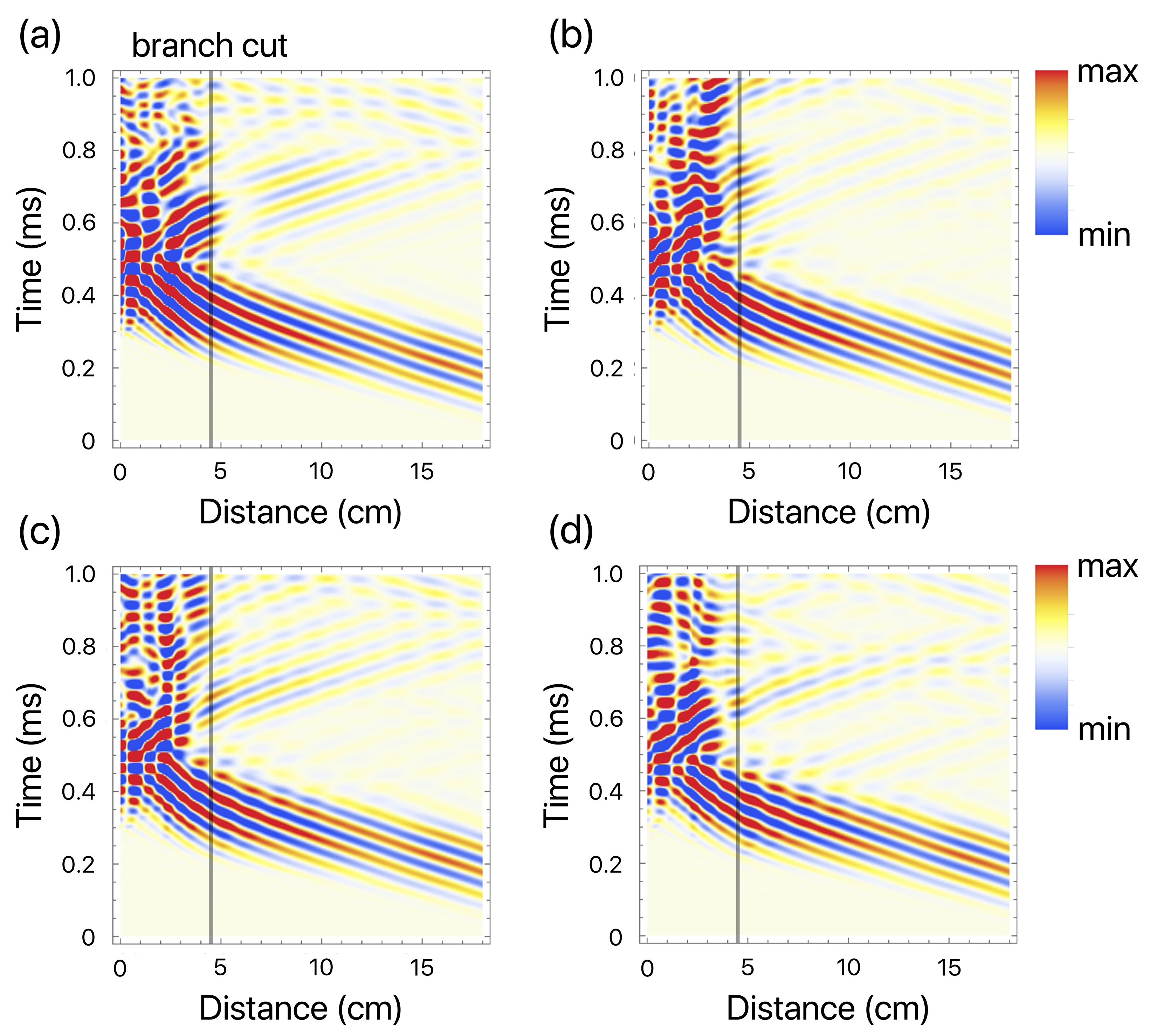}
\caption{Spatiotemporal velocity field maps by varying the thickness of the viscoelastic damping layer; (a) 0.8 mm, (b) 1.6 mm, (c) 2.4 mm, and (d) 3.2 mm.}
\label{fig4}
\end{figure}
Moreover, we investigate spatiotemporal velocity field maps after introducing a complex-valued Young's modulus $E^* = E(1+i\eta)$ where $\eta$ is the loss factor for the viscoelastic damping layer bonded to the branch cut surface. The damping material properties are determined by $\rho=950$ $\mathrm{kg/m^3}$, $E=0.5$ $\mathrm{GPa}$, $\nu=0.45$ and $\eta=0.2$, and the geometrical thickness varies from $0.8$ to $32$ $\mathrm{mm}$ with a step $0.8$ $\mathrm{mm}$ [Figs.~\ref{fig4}(a)-\ref{fig4}(d)]. The viscoelastic damping layer increases the delay by four consecutive pulse widths in time approximately $1$ $\mathrm{ms}$ while lowering reflection from the branch cut. In addition, the velocity of the $A_0$-type Lamb wave at 15 kHz is approximately 507 m/s, thus 1 ms delay is equivalent to a propagation distance of 50.7 cm. It is around 10 times longer than $a$ in the branch cut area and 3 times larger than the whole sample size, making it space-time-efficient. Therefore, we conclude that the elastic landscape with curved elastic continuum is versatile for manipulating waves propagating through solids and revealing such black hole phenomena as condensation and lagging timings inside the branch cut.                                             
\section{\label{sec:level1}Concluding remarks}
Using tactically incorporating a conformal mapping in elastic continuum with curved paradigms, we demonstrate a black hole effect that is particularly noticeable when compared to a Schwarzschild model by Minkowski space-time with an isotropic refractive-index profile. Upon closer inspection of the conformal map, it is discovered that there is a singularity which is achieved by tangible plates with close to zero thickness. Consequently, energy is severely constrained, and timings are delayed inside a branch cut that is roughly analogous to an event horizon in appearance. This behavior is validated numerically and experimentally, and we also carry out control trials under which the thickness is not altered. A deeper comprehension of the elastic counterpart of gravitational events as well as the engineering of an elastic continuum framework for developing a general design strategy in the presence of a singularity would emerge from these findings. We believe that geometries with elastically curved surfaces will find applications in a wide range of categories, including sensing, imaging, vibration isolators, and mechanical-electric energy conversion.

\vspace{1mm}
\section*{Acknowledgements}
This work is financially supported by the POSCO-POSTECH-RIST Convergence Research Center program funded by POSCO, and the National Research Foundation (NRF) Grants (No. NRF-2019R1A2C3003129, CAMM-2019M3A6B3030637, NRF-2019R1A5A8080290) funded by the Ministry of Science and ICT (MSIT) of the Korean government, and the Grant (No. PES4400) from the endowment project of ``Development of smart sensor technology for underwater environment monitoring" funded by Korea Research Institute of Ships \& Ocean engineering (KRISO). D.L. acknowledges the NRF Global Ph.D. fellowship (NRF-2018H1A2A1062053) funded by the Ministry of Education of the Korean government. J.L. acknowledges the funding (Grant No. 16302218, AoE/P-502/20) from the Research Grants Council of Hong Kong.

\vspace{1mm}



\bibliography{BH.bib}

\end{document}